\documentclass[12pt,english]{article}
\usepackage{graphicx, amsmath,amssymb,latexsym,psfrag,epsfig,subfigure,euscript, multirow,color}
\allowdisplaybreaks
\usepackage[T1]{fontenc}
\usepackage[latin9]{inputenc}
\usepackage{slashed}
\usepackage{cite}
\usepackage{graphicx,  amsmath,amssymb,latexsym,psfrag,epsfig,subfigure,euscript, multirow,color,esint}



\usepackage{babel}

\begin{document}

\begin{titlepage}

\begin{flushright}
YITP-SB-11-26\\
UM-DOE/ER/40762-504\\
\end{flushright}

\vspace{0.2cm}
\begin{center}
{\Large\bf Majorana Neutrinos from Inverse Seesaw \\ in Warped Extra Dimension }
\end{center}
\vspace{0.2cm}
\begin{center}
{\sc Chee Sheng Fong$^1$, Rabindra N.\ Mohapatra$^2$ and Ilmo Sung$^2$}\\
\vspace{0.4cm}

$^1$C.N.\ Yang Institute for Theoretical Physics\\
Stony Brook University\\
Stony Brook, New York 11794, U.S.A.
\vspace{0.2cm}

$^2$Maryland Center for Fundamental Physics\\
University of Maryland\\
College Park, Maryland 20742, U.S.A.
\end{center}
\vspace{0.2cm}
\begin{abstract}
\vspace{0.2cm} \noindent
We propose the inverse seesaw mechanism as a way to understand small Majorana masses for neutrinos in
warped extra dimension models with seesaw scale in the TeV range. 
The ultra-small lepton number violation needed  in implementing inverse seesaw mechanism in 4D models is 
explained in this model as a consequence of lepton number breaking occurring on the Planck brane.
We construct realistic models based on this idea that fit 
 observed neutrino oscillation data for both normal and inverted mass patterns.
We compute the corrections to light neutrino masses
from the Kaluza-Klein modes and show that they are small in the parameter range of interest.
Another feature of the model is that the absence of global parity anomaly implies the existence of
 at least one light sterile neutrino with  sterile and active neutrino mixing in the range suggested by
 the LSND and MiniBooNE observations.  
\end{abstract}
\vfil
\end{titlepage}


\section{Introduction} 
Randall-Sundrum\ (RS) hypothesis of the existence of a fifth 
space dimension with a warped metric\ \cite{Randall:1999ee} 
provides an alternative
solution to the gauge hierarchy problem, which is distinct from
the supersymmetric\ (SUSY) approach.  
Different embedding of the Standard Model\ (SM) into warped extra
dimension\ (WED) have been discussed and there has been considerable
attention focused on studying the phenomenological implications and
consistencies of the WED models\ \cite{agashe}.
Understanding
the smallness of neutrino masses in WED models however has been quite non-trivial.  
In contrast, in SUSY
framework, a simple extension of the Minimal Supersymmetric Standard Model (MSSM) 
by the addition of three right-handed neutrinos leads via type I seesaw mechanism\ \cite{seesaw}
 to a set of three light neutrinos. The formula for neutrino masses in this case has
 the form $M_{\nu} \simeq -m_D m_N^{-1} m_D^T$ for $m_D \ll m_N$ where $m_D$ is the Dirac mass
and $m_N$ the Majorana mass of right-handed\ (RH) neutrinos. Since $m_N$ is a new
scale unrelated to the SM gauge group, its value can be much
higher than the weak scale $v_{wk}$ whereas $m_D$ breaks SM gauge
group and is of order of $ v_{wk}$, making $M_\nu$ much smaller than the
known quark and lepton masses. Typical values of the seesaw scale
$m_N$ in Grand Unification Theories (GUTs) are of order $10^{14}\,$GeV. 
A common theme of all seesaw-like solutions to neutrino masses is that neutrinos
are Majorana fermions implying observable lepton number violating
processes. Several ongoing searches for lepton number violating process such
as neutrinoless double beta decay of nuclei are under way to test this hypothesis.

There have been several interesting proposals to understand small neutrino masses in WED
 models\ \cite{Grossman:1999ra,Gherghetta:2000qt,Huber:2003sf,
Gherghetta:2003he,Chen:2005mz,perez,Csaki:2008qq,Agashe:2008fe}.  
A generic prediction of  these models (with the
exception of \cite{Huber:2003sf,perez,Csaki:2008qq}) is that neutrinos are Dirac
fermions so that total lepton number remains a good symmetry of nature
and processes such as neutrinoless double beta decay and $K^+\to \pi^-\mu^+\mu^+$ etc.
that violate lepton number should not
be observed. It has also been argued that this kind of approach
provides a simple way to understand the flavor structure among
neutrinos (a much milder hierarchy for neutrinos compared to
charged leptons)\ \cite{Agashe:2008fe}.  
The models \cite{Huber:2003sf,perez,Csaki:2008qq}
that have Majorana neutrinos use type I seesaw for the purpose so that the seesaw scale
is in the range of $10^{14}\,$GeV or higher and not directly accessible at the LHC.

In this paper, we discuss an alternative class of WED models where neutrinos are Majorana fermions
and obtain their masses from a different mechanism, known in literature as the inverse seesaw
mechanism\ \cite{Mohapatra:1986aw}. Its implementation requires adding two 
gauge singlet chiral fields $N$ and $S$ per family to 
the SM such that they form a pseudo-Dirac pair with  
 mass in the TeV range. The  smallness of neutrino masses is related to
the extent of their ``pseudo-Dirac-ness''  which is governed by a tiny
lepton number breaking mass term for the fields $N, S$ (denoted by $m_{S,N}$). 
The generic mass formula for the neutrino mass matrix is given by
$M_\nu \simeq -m_D \left(m_{SN}\, m_{S}^{-1}\, m_{SN}^T\right)^{-1} m_D^T$ 
with $m_S \ll m_D \ll m_{SN}$ where $m_{SN}$ is the Dirac mass that couples  $N$ and $S$.
Unlike the usual four-dimensional inverse seesaw models, where smallness of $m_S$ 
requires introducing a tiny parameter by hand, 
we show here that in the WED models, one can have this smallness 
dictated by parameters of order one that govern the location of the 5D profile of the
$S$ fields in the bulk. In this sense, the RS framework 
is ideally suited to the implementation of inverse seesaw.
Furthermore, in contrast with the type I embedding in WED, 
the seesaw scale in this case is in the TeV range so that it is accessible at the LHC.

We implement the inverse seesaw mechanism in WED models in this paper and 
present realistic examples that fit current neutrino oscillation data.
An interesting outcome of our model is that it predicts 
the existence of an eV mass sterile neutrino
in a natural manner due to the fact that absence 
of global parity anomaly requires that
there be an even number of singlet $S$-fermions: 
four in our case out of which only three are
required for inverse seesaw, the remaining one will become the light sterile neutrino. 
We note some of the properties
of the sterile neutrino predicted in our model.

This paper is organized as follows: in Sec.\ \ref{sec:typeI_seesaw} we review the
implementation of type I seesaw mechanism in WED\ \cite{Huber:2003sf}. 
In Sec.\ \ref{sec:inverse_seesaw}, we present our model using the inverse seesaw mechanism.
We first illustrate the appearance of light sterile neutrino with a
toy model. We then consider realistic cases and give two examples
 of parameter space 
which reproduce the experimentally measured neutrino mass squared differences and mixing matrix
for normal and inverse neutrino mass hierarchies respectively. 
Then, we study the contribution from higher Kaluza-Klein\ (KK) mode.
In Sec.\ \ref{sec:pheno}, we comment briefly on some phenomenological implication of the model,
in particular the effect on the neutrinoless double beta decay. 


\section{Type I seesaw in warped extra dimension}
\label{sec:typeI_seesaw}

The Randall-Sundrum~(RS) model~\cite{Randall:1999ee} has the warped metric
\begin{eqnarray}
ds^{2} & = & G_{AB}dx^{A}dx^{B}=e^{-2\sigma\left(y\right)}\eta_{\mu\nu}dx^{\mu}dx^{\nu}-dy^{2},
\;\;\;\;\sigma\left(y\right)=k\left|y\right| \, ,
\end{eqnarray}
where $k$ is the AdS curvature, $\eta_{\mu\nu}={\rm diag}\left(1,-1,-1,-1\right)$
and the fifth dimension $-\pi R\leq y\leq\pi R$ is taken to be a $S_{1}/Z_{2}$ orbifold.

As discussed in Ref.\ \cite{Huber:2003sf}, one way to implement type I seesaw in 
WED is to extend the SM by adding three RH neutrinos $N$, 
one for each family and including their Yukawa couplings in 5D. 
The bulk action for this model can be written as follows:
\begin{eqnarray}
S \!& = &\! \int d^{4}x\int_{-\pi R}^{\pi
R}dy\sqrt{G}\left[\overline{N}iE_{a}^{A}\gamma^{a}D_{A}N-m_{DN5}\overline{N}N 
 -\!\!\left( \frac{1}{2}m_{N5}\overline{N}N^{c}
+\lambda_{N5}\overline{\ell}N H +{\rm h.c.} \right)\right]  ,
\label{eq:seesaw_action}
\end{eqnarray}
where $\ell$ and $H$ are respectively the $SU(2)_{L}$ lepton and Higgs doublets
\footnote{To avoid clutter, we have suppressed the family indices of $\ell$ and $N$.}.
In Eq.\ \eqref{eq:seesaw_action}, $a,b,...$ and $A,B,..$ are respectively the flat and curve
indices which run from 0 to 4. We have $\gamma^{a}=\left(\gamma^{\mu},i\gamma^{5}\right)$ 
for $a=0,1,2,3,4$ and the spacetime covariant derivative $D_{A}=\partial_{A}+\omega_{A}$.
From the warped metric, the inverse vielbein is given by 
$E_{a}^{A}={\rm diag}\left(e^{\sigma},e^{\sigma},e^{\sigma},e^{\sigma},1\right)$
while the spin connection is given by 
$\omega_{A}=\left(\frac{1}{2}\sigma'e^{-\sigma}\gamma^{5}\gamma_{\mu},0\right)$
where $\sigma'=d\sigma/dy=k\,{\rm sgn}\left(y\right)$. We then determine
$\sqrt{G}=\sqrt{{\rm det}G_{AB}}=e^{-4\sigma}$. For a spinor $\Psi$, $\Psi^{c}=C\gamma^{0}\Psi^{*}$
is the corresponding charge conjugate spinor with 
$C=i \gamma^{2}\gamma^{0} \gamma^{5}= \gamma^{2}\gamma^{0} \gamma^{4}$ 
such that $\gamma^{a,T}=-C\gamma^{a}C^{-1}$. 

If we assign the lepton number $L$ for both $N,\ell$ as $L=1$
the only term that violates $L$ is Majorana masses $m_{N5}$ 
in the action \eqref{eq:inv_seesaw_action}.
In the standard notation where the warped factor is given by
$e^{-k|y|}$ with the Planck and TeV branes located at $y=0$ and
$y=\pi R$ respectively, the fermion zero modes are given by the 
5D profile $\tilde{f}^{(0)}(y) =  \sqrt{\frac{\pi
kR\left(1-2c_{f}\right)}{e^{\pi
kR(1-2c_{f})}-1}}e^{\left(\frac{1}{2}-c_{f}\right)\sigma}$ with 
Dirac mass parameter $c_f= m_f/k$
 and $\sigma \equiv k |y|$. 
We follow a definition of the profile wave functions 
whose normalization condition does not include any extra warped factor 
(i.e. with respect to flat metric) such that 
\begin{eqnarray}
\frac{1}{2\pi R} \int^{\pi R}_{-\pi R} d y 
\tilde{f}^{(m)} (y) \tilde{f}^{(n)} (y) = \delta_{mn} \, .
\end{eqnarray}
It is clear from this that if $c_f > \frac{1}{2}$, the
profile peaks near the Planck brane whereas if $c_f < \frac{1}{2}$, 
it peaks near the TeV brane. Electroweak precision
constraints demand that the 5D profiles of charged leptons peak near the Planck
brane due to small wave function overlaps with the KK modes\ \cite{agashe}. 
The RH neutrinos being electroweak
singlets do not however have any such constraints. To implement the
seesaw mechanism, lepton number is assumed to be broken at the Planck
brane via the Majorana mass term $m_{N5} = d_N \, \delta (y)$ 
where $d_N$ is a dimensionless number.
The zero mode profile of the RH neutrino is chosen 
to peak near the TeV brane i.e. $c_N < 1/2$.

In order to obtain fermion masses, we need to know the Higgs
doublet profile. We assume that it is localized on the TeV brane. 
To estimate the order of magnitude of the model parameters, we work with
only one generation of fermion. Denoting the Dirac mass parameters for the 
lepton doublet, lepton singlet and RH neutrino respectively 
by $c_{\ell} > 1/2$, $c_{e_R} > 1/2$ and $c_N < 1/2$, 
we can write the effective 4D charged lepton mass $m_\ell$, 
Dirac mass for the neutrinos $m_D$ and the Majorana mass $m_N$ 
for the RH neutrinos as:
\begin{eqnarray}
m_\ell ~&\sim& k\times e^{-k\pi R(c_\ell+c_{e_R})}\, , \nonumber \\
m_D~&\sim&~k\times e^{-k\pi R(c_\ell+\frac{1}{2})}\, , \nonumber \\
m_N~&\sim&~ k\times e^{k\pi R(2c_N-1)} \, ,
\end{eqnarray} 
which leads to light neutrino mass 
\begin{equation}
m_\nu \sim k\times e^{-2k\pi R(c_\ell+c_N)} \, .
\label{eq:ss_neutrino_mass}
\end{equation}
Here all 5D dimensionless couplings are assumed to be unity. 
With $k\pi R \sim 37$ and $k = 2.4 \times 10^{18}\,{\rm GeV}$,
for example, in order to get the right charged lepton masses, 
we choose $c_{\ell_\alpha}=0.65$ ($\alpha = e, \mu, \tau$), 
$c_{e_R}~=~0.78$, $c_{\mu_R}=0.61$ and $c_{\tau_R}~=~0.53$. 
Since $e^{-74}\sim 7\times 10^{-33}$, to get neutrino masses 
$m_\nu \lesssim 1\,{\rm eV}$, we will have $c_N \gtrsim 0.2$.  
Here the smallness of the light neutrino mass is attributed to
a large $m_N$ as in the usual seesaw. 
Since we place the hierarchy in $c_{\alpha_R}$ and fix the 
$c_{\ell_\alpha}$ to be the same for all flavors, we will get
a non-hierarchical (anarchical) neutrino mass matrix if $c_N$ is non-hierarchical.
On the other hand, if we fix $c_{\alpha_R}$ while having hierarchical
$c_{\ell_\alpha}$, $c_N$ should also be hierarchical in order to get
an anarchical neutrino mass matrix.


\section{Inverse seesaw in warped extra dimension}
\label{sec:inverse_seesaw}

As noted earlier, to implement the inverse seesaw mechanism\ \cite{Mohapatra:1986aw} 
in 4D models, two types of chiral gauge singlet fermions are needed. 
As before we denote by $N$ the RH neutrinos used for seesaw mechanism discussed above 
and by $S$ the extra singlet fermion fields. They form a pseudo-Dirac pair with
 splitting given by a tiny parameter that breaks the lepton number. 
The smallness of this parameter is chosen by hand in the 4D case.
We follow this strategy closely in the discussion of WED embedding of inverse seesaw.
The bulk action for inverse seesaw in WED can be written as follows:
\begin{eqnarray}
S & = & \int d^{4}x\int_{-\pi R}^{\pi R}dy\sqrt{G}
\Bigg[\overline{N}iE_{a}^{A}\gamma^{a}D_{A}N-m_{DN5}\overline{N}N
 +\overline{S}iE_{a}^{A}\gamma^{a}D_{A}S-m_{DS5}\overline{S}S\nonumber \\
 &  & -\left( \frac{1}{2}m_{N5}\overline{N}N^{c}
+\frac{1}{2}m_{S5}\overline{S}S^{c}+\frac{1}{2}m_{SN5}\overline{S}N^{c}
+\lambda_{N5}\overline{\ell}N H+\lambda_{S5}\overline{\ell}S^{c} H
+{\rm h.c.}
\right)\Bigg] \, .
\label{eq:inv_seesaw_action}
\end{eqnarray}
If we assign the lepton number $L$ for $N,S$ respectively as $L=1,-1$,  
the only fermion bilinears which violate $L$ are the Majorana masses
$m_{N5}$ and $m_{S5}$ in the action \eqref{eq:inv_seesaw_action}. 

This above action could also arise from an exact gauge symmetry 
such as $U(1)_{B-L}$\ (with the usual definition of quantum numbers) 
after spontaneous symmetry breaking by Higgs field that transforms as
$B-L =+1$.  The $S$ field is $B-L$ neutral and therefore 
its Majorana mass term $m_{S5}$ which is allowed by $B-L$ breaks the global symmetry $L$\ (defined above) which persists
in gauged $B-L$ version.  
The $B-L$ model implies that $m_{N5}=0$; however,
since $m_{N5}$ does not play a role in the neutrino masses and mixing, 
the final results  derived in the model without $B-L$ symmetry and presented below remain unchanged. 

Note that in 4D, a five dimensional field splits into two chiral pairs 
and only one chirality remains as a zero mode. So in our model, in 4D, 
only the left chirality of the lepton doublet $\ell$ 
and the right chiralities of $S$ and $N$ survive as zero modes. 

In odd space-time dimension (i.e. five), the action \eqref{eq:inv_seesaw_action} 
contains parity anomaly if the total number of 
bulk fermions that couple to gauge and gravity fields
is odd\ \cite{Redlich:1983kn,Callan:1984sa}. In warped type I seesaw as discussed 
in Sec.\ \ref{sec:typeI_seesaw} where the lepton doublets also propagate in the extra dimension, 
cancellation of the parity anomaly naturally requires three generations 
of RH neutrinos $N$. However, in the warped inverse seesaw, 
in order not to reintroduce parity anomaly, 
we have to add an even number of singlet Dirac fermions $S$.
The minimal number of $S$ fields required to obtain three active light neutrino 
is three. Since we cannot have odd number of $S$, the minimal number has
to be four. Thus, after the three of the four $S$-fields pair up 
with the three $N$ fields to make the three pseudo-Dirac fermions,
we are left with an extra $S$ field which in the end becomes the 
sterile neutrino with mass in the eV range.

We again assume that Higgs doublet is localized on the TeV brane 
and that the $L$-violating Majorana masses are confined to the Planck brane
i.e. $m_{N5} = d_N \, \delta(y),\, m_{S5} = d_S \, \delta(y)$ with $d_N,\,d_S$
dimensionless numbers. 
For simplicity, 
we further assume that $m_{SN5} = d_{SN} \, k$
with $d_{SN}$ dimensionless number and ignore any possible boundary masses. 
To estimate the order of magnitude of the dimensionless parameters 
that characterize the model, we consider only one generation for all fermions. 
Assuming the 5D location of the fields to be 
$c_\ell > 1/2$, $c_{e_R} > 1/2$, $c_N < 1/2$ and $c_S < 1/2$,
we find the 4D effective masses to be 
\begin{eqnarray}
m_D~&\sim&~ k\times e^{-k\pi R(c_\ell+\frac{1}{2})}\, , \nonumber \\
m_N ~&\sim&~ k \times e^{\pi k R(2c_N-1)} \, ,\nonumber \\ 
m_S ~&\sim&~ k \times e^{\pi k R(2c_S-1)} \, , \nonumber \\
m_{SN}~&\sim&~ k \times e^{-k \pi R}  \, . 
\end{eqnarray} 
In the equation for $m_{SN}$ we have also assumed 
that $ c_N + c_S \leq 0$ and we will see that this is indeed what 
we need for the inverse seesaw mechanism to work.
This leads to the effective light neutrino mass
\begin{equation}
m_{\nu} \sim k \times e^{-2 \pi k R (c_\ell  -c_S)} \, .
\label{eq:invss_neutrino_mass}
\end{equation}
In contrast to Eq.~\eqref{eq:ss_neutrino_mass}, here
the smallness of the light neutrino mass is attributed to small $m_S$.
For example, if we take $c_\ell = 0.65$ and $c_S \lesssim -0.2$,
we have $m_\nu \lesssim 1\,{\rm eV}$ with the seesaw scale 
$m_{SN} \sim \cal{O}$(TeV). Also, if we assume all $c_\ell$ and $c_S$ to be 
of same order for all generation, we get a neutrino 
mass matrix with non-hierarchical pattern.
It is interesting that the final neutrino mass formula 
is independent of the precise 5D profile of the $N$ fields.


\subsection{The appearance of sterile neutrino(s)}
\label{sec:sterile}

To understand the appearance of the sterile neutrino(s), 
let us consider a toy model with one generation of 
$\ell$ and $N$ and two generations of $S$ called $S_1, S_2$.
In this case, since only one S is needed for the inverse seesaw, the additional one
would result in a sterile neutrino.
In this model, the mass matrix in the basis $(\nu,N,S_1,S_2)$
is given by
\begin{eqnarray}
M & = & \left(\begin{array}{cccc}
0 & m_D & 0 & 0 \\
m_D & m_N & m_{SN_1} & m_{SN_2} \\
0 & m_{SN_1} & m_{S_{11}} & m_{S_{12}} \\
0 & m_{SN_2} & m_{S_{12}} & m_{S_{22}} 
\end{array}\right) \, .
\label{eq:model_112}
\end{eqnarray}

For simplicity, we assume that $m_{SN_1} = m_{SN_2} = m_{SN}$,
$m_{S_{11}} = m_{S_{22}} = m_S$ and $m_N = m_{S_{12}} = 0$.
Assuming $m_S \ll m_D \ll m_{SN}$, we can diagonalize matrix
\eqref{eq:model_112} and obtain two heavy 
and two light states with their respective masses given by
\begin{eqnarray}
m_{\rm heavy} &\simeq& \pm \sqrt{2m_{SN}^2 + m_D^2} 
+ m_S\frac{m_{SN}^2}{2 m_{SN}^2 + m_D^2} \, , \label{eq:m_heavy}\\
m_{\rm light} &\simeq& m_S,\;\; m_S\frac{m_D^2}{2m_{SN}^2 + m_D^2} \, .
\label{eq:m_light}
\end{eqnarray}
The two heavy states mix with the light neutrino with $\sim m_D/m_{SN}$ 
and can be named the heavy RH neutrinos.
The light state with mass $m_S$ can be identified as sterile
neutrino while the other is the light active neutrino.
Hence, we obtain an interesting relation between the mass
of active and sterile neutrinos as follows
\begin{equation}
m_{\rm active} \simeq m_{\rm sterile}\, 
\frac{m_D^2}{2m_{SN}^2 + m_D^2} \, .
\label{eq:mac_mst}
\end{equation}
For instance, to have an active neutrino with mass $m_{\rm active}\sim$ 0.05 eV
and a sterile neutrino with mass $m_{\rm sterile}\sim$ 1 eV, we would require 
a hierarchy between $m_D$ and $m_{SN}$ to be $m_D/m_{SN} \sim 0.22$.
On the other hand, if we want $m_{\rm sterile}\sim$ 1 keV
which could be a potential dark matter candidate, it would require 
$m_D/m_{SN} \sim 0.007$. 

It should be pointed out that although the result above 
is obtain by assuming $m_{S_{12}}$ to be vanishing,
barring any accidental cancellation, it holds in general even if $m_{S_{12}}$ 
is of the order of the diagonal elements $m_{S_{11}}$ and $m_{S_{22}}$.
The result will also hold if $m_N$ is non-vanishing as
long as $m_N \ll m_{SN}$. In order to obtain more than one sterile neutrino,
we can extend the number of $S$ in the model. 


\subsection{Neutrino mixing in warped inverse seesaw}
\label{sec:mixing}

We will now present a realistic warped inverse seesaw model with 
three $N$ and four $S$ fields to explore the detailed 
neutrino mixing and mass hierarchy pattern.
We first ignore the contributions of the KK modes 
which will be discussed in a subsequent section, where we will show 
under what conditions their effects can be safely ignored.
Considering for now only the zero modes, we have 
the leading $10 \times 10$ neutrino mass matrix in the basis
$(\nu,N,S)$, which is given as follows
\begin{eqnarray}
M & = & \left(\begin{array}{ccc}
0 & m_D & 0 \\
m_D^T & m_N & m_{SN} \\
0 & m_{SN}^T & m_S 
\end{array}\right) \, .
\label{eq:invss_mass_matrix}
\end{eqnarray}
where $m_D$, $m_{SN}$ and $m_S$ are respectively 
$3\times3$, $3\times 4$ and $4\times 4$ matrices.
Assuming $m_S,m_N \ll m_D \ll m_{SN}$, we can block diagonalize the 
mass matrix above and obtain the light neutrino mass matrix to be
\begin{equation}
M_\nu \simeq -m_D \left(m_{SN}\,m_S^{-1}\,m_{SN}^T\right)^{-1} m_D^T \, .
\label{eq:invss_light_neutrino}
\end{equation}
Note that  $m_N$ does not appear in the light neutrino masses. It only affects
the mass splitting of the pseudo-Dirac pair $(N,S)$. 


\subsection{Examples I: Normal hierarchy (NH) mass spectrum}
\label{sec:nor_eg}

In this section, we address the issue of neutrino mixing. 
We will search for the parameter space which reproduces the neutrino mixing matrix 
for the normal hierarchy spectrum ($m_{\nu_3} > m_{\nu_2} > m_{\nu_1}$) 
while having anarchical pattern for $m_D$ and $m_{SN}$. Notice
that $m_{SN}$ is naturally anarchical. However, as far as $m_D$ is 
concerned, unlike the RH charged leptons
since the $N$ is located closer to the TeV brane, whether it is hierarchical or not depends
on the profiles of the left-handed charged leptons.
By fixing the values of left-handed lepton doublets and attributing the hierarchy to 
the RH singlet charged leptons, we can have an anarchical $m_D$. 
For example, we have chosen the bulk mass 
parameters for charged leptons as follows: $c_{\ell_e} = c_{\ell_\mu} = c_{\ell_\tau} = 0.65$, 
$c_{e_R} = 0.7770, c_{\mu_R} = 0.6099, c_{\tau_R} = 0.5271 $. 
This fits the charged lepton mass spectrum. We then choose
$c_{N_1} = c_{N_2} = c_{N_3} = -0.340$ and
$c_{S_1} = -0.338, c_{S_2} = -0.366, c_{S_3} = -0.358, c_{S_4} = -0.377$,
and all the dimensionless couplings having the values in the range $[0.1,1.0]$ .
We then  obtain the following mass matrices\ (in GeV)

\begin{eqnarray}
m_{D} & = & \left(\begin{array}{ccc}
 2.763 & 6.029 & 15.826 \\
 9.294 & 5.778 & 12.058 \\
 12.540 & 16.077 & 7.285
\end{array}\right) \, ,
\label{eq:mD_anar_nor}
\end{eqnarray}

\begin{eqnarray}
m_{SN} & = & \left(\begin{array}{cccc}
 172.342 & 191.492 & 138.988 & 141.208 \\
 177.903 & 222.665 & 134.505 & 264.764 \\
 82.1093 & 276.105 & 347.918 & 269.177
\end{array}\right) \, ,
\label{eq:mSN_anar_nor}
\end{eqnarray}

\begin{eqnarray}
m_{S} & = & \left(\begin{array}{cccc}
 5.8522 & 3.2171 & 2.5562 & 1.8856 \\
 3.2171 & 0.7979 & 2.5724 & 1.5791 \\
 2.5562 & 2.5724 & 3.8971 & 0.8128 \\
 1.8856 & 1.5791 & 0.8128 & 1.3833
\end{array}\right)\times 10^{-9} \, ,
\label{eq:mS_anar_nor}
\end{eqnarray}

\begin{eqnarray}
m_{N} & = & \left(\begin{array}{ccc}
 0.4072 & 1.018 & 0.6108 \\
 1.018 & 0.8143 & 1.222 \\
 0.6108 & 1.222 & 1.018
\end{array}\right)\times 10^{-9} \, .
\label{eq:mN_anar_nor}
\end{eqnarray}

From the matrices above, we obtain the light neutrino mixing matrix 
(by diagonalizing the $10 \times 10$ neutrino mass matrix) as follows
\begin{eqnarray}
U_\nu^{nor} & = & \left(\begin{array}{cccc}
 -0.8517 & 0.5122 & 0.0962 & -0.0135 \\
 0.3183 & 0.6593 & -0.6694 & 0.1104 \\
 -0.4136 & -0.5468 & -0.7182 & 0.1005 \\
 0.0466 & 0.0633 & 0.1638 & 0.9887
\end{array}\right) \, ,
\label{eq:neutrino_mixing_nor}
\end{eqnarray}
where the last row and column correspond to the mixing with light sterile neutrino.
The top left $3\times 3$ submatrix of matrix \eqref{eq:neutrino_mixing_nor} 
corresponds to the mixing between active neutrinos and is in good
agreement with the one obtained from the approximate formula 
Eq.\ \eqref{eq:invss_light_neutrino}\footnote{In this work, we will only use 
the exact mixing matrix obtained from diagonalizing $10 \times 10$ neutrino mass matrix.}.
The active neutrino mixing matrix \eqref{eq:neutrino_mixing_nor} is in
good agreement with the experimentally measured values~\cite{GonzalezGarcia:2010er}
\begin{eqnarray}
U_\nu^{exp} & = & \left(\begin{array}{ccc}
-0.8212 & 0.5623 & 0.0976 \\ 
0.3598 & 0.6429 & -0.6762 \\
-0.4429 & -0.5202 & -0.7302 
\end{array}\right) \, .
\label{eq:neutrino_mixing_exp}
\end{eqnarray}

The masses of three light active neutrinos are 
$(\nu_1,\nu_2,\nu_3) = (0.00172, 0.00885, 0.0516)\,{\rm eV}$. 
On the other hand, the sterile neutrino
has a mass of $0.848\,{\rm eV}$ which could potentially explain 
the anomaly in LSND and MiniBooNE\ \cite{Aguilar:2001ty,AguilarArevalo:2010wv}.
From the above, we can determine the mass squared differences 
of the active neutrinos
\begin{eqnarray}
\Delta m_{21}^2 & = & 7.54 \times 10^{-5}\,{\rm eV}^2 \, , \nonumber \\
\Delta m_{31}^2 & = & 2.66 \times 10^{-3}\,{\rm eV}^2 \, ,
\end{eqnarray}
which are within $1 \sigma$ of the experimental values.

With the existence of an extra light sterile neutrino with 
significant mixing with the active neutrinos, we have to
check if this could be consistent with the experimentally
determined number of species of neutrinos. 
In the SM, the difference of the total width of the $Z^0$ boson 
and the width for the decay into all visible channels is 
attributed only to the light neutrinos that couple to the $Z^0$ boson.
The was determined very precisely from LEP data to be 
$N_\nu = 2.9841 \pm 0.0083$~\cite{Nakamura:2010zzi}.
In our example above, we can calculate the correction due to the
existence of the light sterile neutrino which mixes with 
active neutrinos~\cite{Bilenky:1990tm}
\begin{equation}
N_\nu = \sum_{i,j=1}^4 
\left|\sum_{\alpha=e,\mu,\tau} 
U_{\alpha i}^* U_{\alpha j}\right|^2 = 2.979 \, ,
\label{eq:zwidth_num_nu}
\end{equation}
where we have ignored the neutrino masses since $m_{1,2,3,4} \ll M_{Z^0}$.

Finally we can also write down $4\times 4$ neutrino mass matrix 
including three active and one light sterile neutrinos as follows\ (in GeV)
\begin{eqnarray}
M_{\nu}^{nor} & = &
\left(
\begin{array}{cccc}
 0.003890 & 0.0004645 & -0.004287 & 0.01234 \\
 0.0004645 & 0.01681 & 0.01198 & -0.09776 \\
 -0.004287 & 0.01198 & 0.02098 & -0.09063 \\
 0.01234 & -0.09776 & -0.09063 & -0.8271
\end{array}
\right) \times 10^{-9} \, .
\end{eqnarray}

Since the experiment could not determine the sign of $\Delta m_{31}^2$, 
we consider below the possibility of inverted  neutrino mass spectrum  i.e.
$m_{\nu_2} > m_{\nu_1} > m_{\nu_3}$ .


\subsection{Examples II: Inverted hierarchy (IH) mass spectrum}
\label{sec:inv_eg}

We again choose anarchical $m_D$ and $m_{SN}$. 
The 5D parameters in this case are different. 
For example, we have chosen the bulk mass 
parameters as follows: $c_{\ell_e} = c_{\ell_\mu} = c_{\ell_\tau} = 0.65$, 
$c_{e_R} = 0.7770, c_{\mu_R} = 0.6099, c_{\tau_R} = 0.5271 $,
$c_{N_1} = c_{N_2} = c_{N_3} = -0.360$,
$c_{S_1} = -0.3869, c_{S_2} =-0.353, c_{S_3} = -0.3029, c_{S_4} =-0.343$,
and all the dimensionless couplings having the values in the range $[0.1,1.0]$ 
and we obtain the following mass matrices (in GeV)

\begin{eqnarray}
m_{D} & = & \left(\begin{array}{ccc}
 3.1553 & 5.7058 & 7.4499 \\
 8.1511 & 3.9266 & 2.7872 \\
 3.8560 & 5.5772 & 0.3675
\end{array}\right) \, ,
\label{eq:mD_anar_inv}
\end{eqnarray}

\begin{eqnarray}
m_{SN} & = & \left(\begin{array}{cccc}
 105.405 & 101.936 & 107.203 & 70.534 \\
 85.528 & 110.527 & 73.175 & 133.853 \\
 32.219 & 150.114 & 176.516 & 141.532
\end{array}\right) \, ,
\label{eq:mSN_anar_inv}
\end{eqnarray}

\begin{eqnarray}
m_{S} & = & \left(\begin{array}{cccc}
 0.2383 & 0.9876 & 4.0775 & 1.5672 \\
 0.9876 & 2.0009 & 29.796 & 8.5212 \\
 4.0775 & 29.796 & 214.205 & 20.529 \\
 1.5672 & 8.5212 & 20.529 & 16.493
\end{array}\right)\times 10^{-9} \, ,
\label{eq:mS_anar_inv}
\end{eqnarray}

\begin{eqnarray}
m_{N} & = & \left(\begin{array}{ccc}
 0.09489 & 0.2372 & 0.1423 \\
 0.2372 & 0.1898 & 0.2847 \\
 0.1423 & 0.2847 & 0.2372
\end{array}\right)\times 10^{-9} \, .
\label{eq:mN_anar_inv}
\end{eqnarray}

From the matrices above, we obtain the light neutrino mixing matrix 
\begin{eqnarray}
U_\nu^{inv} & = & \left(\begin{array}{cccc}
 -0.8131 & -0.5714 & 0.09739 & -0.0358 \\
 0.3440 & -0.6076 & -0.7062 & -0.0866 \\
 -0.4635 & 0.5353 & -0.6964 & 0.1003 \\
 -0.0755 & 0.1336 & 0.0834 & -0.9905
\end{array}\right) \, ,
\label{eq:neutrino_mixing_inv}
\end{eqnarray}
where again the last row and column correspond to the mixing with light sterile neutrino.
As before, the top left $3\times 3$ submatrix of matrix \eqref{eq:neutrino_mixing_inv} 
corresponds to the mixing between active neutrinos 
and is in good agreement with the measured value in 
Eq.\ \eqref{eq:neutrino_mixing_exp}.~\footnote{The sign differences in the 
second column of Eq.\ \eqref{eq:neutrino_mixing_inv} 
can be accounted for by changing the Majorana phases accordingly.}

The masses of three light active neutrinos are 
$(\nu_1,\nu_2,\nu_3) = (0.0471, 0.0479, 0.000454)\,{\rm eV}$.  
On the other hand, the sterile neutrino has a mass of $16.9\,{\rm eV}$ 
which could be too large to explain the anomaly in LSND and MiniBooNE.
From the above, we can determine the mass squared differences 
of the active neutrinos
\begin{eqnarray}
\Delta m_{21}^2 & = & 7.74 \times 10^{-5}\,{\rm eV}^2 \, , \nonumber \\
\Delta m_{31}^2 & = & - 2.22 \times 10^{-3}\,{\rm eV}^2 \, ,
\end{eqnarray}
which are within 1$\sigma$ of the experimental values for the inverted mass spectrum.
In this example, we determine $N_\nu = 2.9767$ from Eq.\ \eqref{eq:zwidth_num_nu}.

We can also write down $4\times 4$ neutrino mass matrix 
including three active and one light sterile neutrinos as follows\ (in GeV)
\begin{eqnarray}
M_{\nu}^{inv} & = &
\left(
\begin{array}{cccc}
 0.03718 & 0.02257 & -0.02833 & 0.6064 \\
 0.02257 & 0.1148 & -0.1385 & 1.4528 \\
 -0.02833 & -0.1385 & 0.1770 & -1.6816 \\
 0.6064 & 1.4528 & -1.6816 & 16.5961
\end{array}
\right) \times 10^{-9} \, .
\end{eqnarray}


\subsection{Contributions from Kaluza-Klein modes}
\label{sec:kk-modes}

In this section we would like to estimate the contributions
from KK modes. For simplicity, we would consider single generation
for each $N$, $S$ and $\ell$ fields. 
We KK decompose the 5D fermionic fields as
\begin{equation}
\Psi_{L,R}(x^\mu,y) = \frac{e^{2\sigma}}{\sqrt{2\pi R}} 
\sum_{n=0}^{\infty} \Psi^{(n)}_{L,R}(x^\mu) \widetilde\Psi^{(n)}_{L,R}(y) \, ,
\label{eq:kk-decom}
\end{equation}
where $\Psi_{L,R} = \frac{1}{2}(1\mp \gamma_5)\Psi$. 
For $N$ and $S$, we choose $N_R$ and $S_R$ to be even under $Z_2$
while for $\ell$, we choose $\ell_L$ to be even.
As before, we will restrict $H$ to be strictly confined to the TeV brane 
with $H(y) = k \, \delta(y-\pi R)$ and the Majorana masses to be strictly 
confined to the Planck brane i.e. $m_{N5} = d_N \, \delta (y)$ and 
$m_{S5} = d_S \, \delta (y)$. Similarly, we also assume
$m_{SN5} = d_{SN} k$.

Substituting Eq.\ \eqref{eq:kk-decom} for $N$, $S$ and $\ell$ fields 
respectively into the action \eqref{eq:inv_seesaw_action},
we can write down the mass matrix in the basis of 
$\left(\nu_{L}^{(0)},N_{R}^{\left(0\right)},S_{R}^{\left(0\right)},
\nu_{L}^{(1)},\nu_{R}^{(1)},
N_{L}^{\left(1\right)},N_{R}^{\left(1\right)},
S_{L}^{\left(1\right)},S_{R}^{\left(1\right)},...\right)$
as follows
\begin{eqnarray}
\!\!\!\! M^{kk} & = & \left(\begin{array}{cccccccccc}
0 & m_{D}^{(00)} & 0 & 0 & 0 & 0 & 
m_{D}^{(01)} & 0 & 0 & ...\\
m_{D}^{(00)} & m_{N_{R}}^{(00)} & m_{SN_{R}}^{(00)} & m_{D}^{(10)} & 0  &
0 & m_{N_R}^{(01)} & 0 & m_{SN_R}^{(10)} & ...\\
0 & m_{SN_{R}}^{(00)} & m_{S_R}^{(00)} & 0 & 0 &
0 & m_{SN_{R}}^{(01)} & 0 & m_{S_{R}}^{(01)} & ...\\
0 & m_{D}^{(10)} & 0 & 0 & m_\nu^{(1)} &
0 & m_{D}^{(11)} & 0 & 0 & ...\\
0 & 0 & 0& m_\nu^{(1)} & 0 &
0 & 0 & 0 & 0 & ...\\
0 & 0 & 0 & 0 & 0 &
0 & m_{N}^{(1)} & m_{SN_{L}}^{(11)} & 0 & ...\\
m_{D}^{(01)} & m_{N_{R}}^{(01)} & m_{SN_{R}}^{(01)} 
& m_{D}^{(11)} & 0 & m_{N}^{(1)} & m_{N_{R}}^{(11)} 
& 0 & m_{SN_{R}}^{(11)} & ...\\
0 & 0 & 0 & 0 & 0 &
m_{SN_{L}}^{(11)} & 0 & 0 & m_{S}^{(1)} & ...\\
0 & m_{SN_{R}}^{(10)} & m_{S_{R}}^{(01)} & 0 & 0 &
0 & m_{SN_{R}}^{(11)} & m_{S}^{(1)} & m_{S_{R}}^{(11)} & ...\\
... & ... & ... & ... & ... & ... & ... & ... & ... & ...
\end{array}\right)\, ,
\label{eq:mass_matrix}
\end{eqnarray}
where $m_S^{(1)},\, m_N^{(1)}$ and $m_\nu^{(1)}$ are respectively 
the first KK masses of $S,\, N$ and $\nu$ and
\begin{eqnarray}
m_{D}^{(mn)} & = & e^{-k \pi R} 
\lambda_{N4}\, k\, \widetilde \nu_{L}^{(m)} (\pi R)
\widetilde N_{R}^{(n)} (\pi R) \, , \nonumber \\
m_{N_{R}}^{(mn)} & = & 
\frac{d_N}{2\pi R} \widetilde N_{R}^{(m)} (0)
\widetilde N_{R}^{(n)} (0) \, , \nonumber \\
m_{S_{R}}^{(mn)} & = & 
\frac{d_S}{2\pi R} \widetilde S_{R}^{(m)} (0)
\widetilde S_{R}^{(n)} (0) \, , \nonumber \\
m_{SN_{L,R}}^{(mn)} & = &  
\int^{\pi R}_{-\pi R} \frac{dy}{2\pi R} \, e^{-k |y|} \,  
d_{SN}\, k \,\widetilde S_{L,R}^{(m)} (y)
\widetilde N_{L,R}^{(n)} (y) \, .
\label{eq:mass_elements_sim}
\end{eqnarray}
In the first equation of Eqs.\ \eqref{eq:mass_elements_sim},
$\lambda_{N4} = \frac{\lambda_{N5}}{2\pi R}$ is dimensionless.

Assuming $m_{N_{R}}^{(mn)},m_{S_{R}}^{(mn)} \ll 
m_D^{(mn)} \ll m_{SN_{L,R}}^{(mn)} \ll \mbox{KK masses}$,
we obtain the light neutrino mass roughly as
\begin{eqnarray}
m_{\nu} & \sim & m_{S_{R}}^{(00)}
\frac{\left(m_D^{(00)}\right)^{2}}{\left(m_{SN_{R}}^{(00)}\right)^{2}}
\left[1+\mathcal{O}\left(\epsilon^{2}\right)\right] \, ,
\label{eq:light_neutrino_mass}
\end{eqnarray}
where $\epsilon \sim \frac{m_{SN_{L,R}}^{(mn)}}{\mbox{KK masses}}$.
Notice that at leading order, the light neutrino mass is given 
by the inverse seesaw relation and the contributions from 
KK modes are suppressed as long as 
$m_{SN_{L,R}}^{(mn)}$ is less than KK masses.

\begin{figure}[fhptb]
\centering
\includegraphics[scale=0.7]{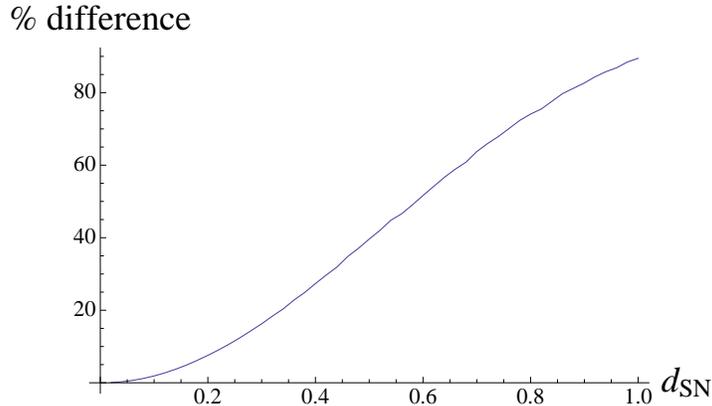}
\caption[]{The contributions from the first KK modes 
to the light neutrino mass as a function of $d_{SN} = m_{SN5}/k$.}
\label{fig:kk}
\end{figure}

Including the contributions from the first KK modes by considering
the entire $9 \times 9$ mass matrix in Eq.\ (\ref{eq:mass_matrix}), 
in Figure \ref{fig:kk}, we numerically plotted the 
contributions of the first KK modes to the light 
neutrino mass as a function of $d_{SN} = m_{SN5}/k$. 
As long as we keep $d_{SN} \lesssim 0.3$ 
(as we did in the examples in Secs.~\ref{sec:nor_eg} and \ref{sec:inv_eg}),
the corrections from the first KK modes are not more than 20 \%.
Hence, we expect the contributions from higher KK modes to be negligible.

\section{Comments }
\label{sec:pheno}
A few comments  are now in order about our model.

\noindent (i) In the above discussion,  we have added two kinds of lepton number breaking terms on the
Planck brane and assumed that these are the only sources of lepton number violation in our model
 i.e. Majorana mass terms of type $NN$ and $SS$. However, 
we could just keep only the second of the two terms, in which case
in Eq.\ (\ref{eq:invss_mass_matrix}) depicting the inverse seesaw matrix for the zero modes, 
the term $m_N$ will absent. Similarly in the discussion of KK mode contributions,
all Majorana terms involving $N_L, N_R$ (i.e. $m^{(mn)}_{N_{L,R}}$) will be absent. 
This makes it easier to estimate the KK contributions to the zero mode
mass and it confirms our result that they are indeed small.  
Such a situation can be guaranteed by adding an extra $B-L$ gauge symmetry 
into the theory under which $S$ is a singlet but $N$ field is not. 
The $m_{SN}$ is then generated by a Higgs field which breaks the 
$B-L$ symmetry by one unit. Since $m_N$ and all Majorana masses involving the higher 
KK modes of the $N$ field break $B-L$ by two units, they will be absent.

\noindent (ii) A comment on the phenomenology of our model: 
A key feature of the model is the presence of 
a light sterile neutrino, which arises because 
of the need to guarantee freedom from parity anomalies as noted.
Since the number of $S$ field $N_S$ we can add to the model has to be 
even, the prediction of this model is that we will have an odd number
of sterile neutrinos $N_{\rm sterile} = N_S - 3$ where the 3
is the number of family in the SM. 
The sterile neutrino will
contribute to neutrinoless 
double beta decay due to its mixing with $\nu_e$; however
the effective neutrino mass due to this contribution 
remains in the 3 meV range due to small mixing and eV range sterile mass.
This remains far below the reach of the current double beta decay search.
This sterile neutrino could also provide a way to understand the recent 
reactor anomalies as well as the MiniBooNE and LSND
results~\cite{Giunti:2011gz}. However at LSND and MiniBooNE, 
it will predict the same oscillation effect for both  neutrinos and anti-neutrinos. 
The sterile neutrino will contribute an extra neutrino 
species in the analysis of Big Bang nucleosynthesis (BBN). This is consistent
with current analyses of the BBN as well as cosmological 
structure formation and WMAP data~\cite{hannestad}.

\noindent(iii) 
The scenario outlined here leads to a $\theta_{13}\simeq 0.096$ 
for the NH and $0.097$ for the IH case. This is however 
not a prediction since the Dirac neutrino Yukawa coupling, the
lepton number violating masses $m_{S}$, $m_{N}$
as well as the $m_{SN}$ matrices all involve free parameters.

\noindent (iv) The specific model discussed here predicts RH neutrinos 
with masses of order 100 GeV which are therefore 
accessible at the LHC via their mixing with the left-handed neutrino.
LHC signals for such Dirac neutrinos have been studied in Ref.\ \cite{saavedra}.
Their primary decay signal is the three lepton plus missing energy in $pp$ collisions.  Furthermore,
an interesting possibility is the KK excited mode of
the  electron, if in the TeV range could decay to the RH neutrino and the $W$. 
Since the dominant decay mode of the RH neutrino is 
to two leptons plus missing energy ($N\to \ell^+\ell^- \nu$),
there could be exotic final states such as $\ell^\pm\ell^\mp \ell^- \nu$.

\noindent(v) The TeV scale particle spectrum in the model is similar 
to an SO(10) model with inverse seesaw discussed in the literature. 
Extrapolating the discussion of that model~\cite{blanchet}, it appears very likely  
that it will provide a satisfactory framework for  realization 
of resonant leptogenesis idea to understand the origin of matter.


\section{Conclusions}
\label{sec:conclusion}

In summary, we have presented a new way to understand the small neutrino masses
by embedding the inverse seesaw mechanism into the warped extra dimension models.  
In the four dimensional implementation of inverse seesaw, 
a small lepton number violating mass term needs to be
put in by hand (of order of or less than a keV) to get sub-eV scale active neutrino masses. 
In the WED framework on the other hand, locating the 
lepton number violating mass terms on the Planck brane
provides a simple way to understand this smallness without any fine-tuning of parameters.
This model differs from the type I seesaw in WED by the 
presence of sub-TeV scale right-handed neutrinos which may be accessible at the LHC.
An interesting prediction of the model is an eV scale sterile 
neutrino which arises from the requirement of cancellation of
parity anomaly in odd number of dimensions. Its small mass is again connected to the small
parameter in the inverse seesaw and the lepton number breaking in the Planck brane.
We have also worked out numerical examples which give active neutrino masses and mixing 
in accord with observations for both normal and inverted hierarchy cases, 
showing that such models can indeed provide realistic description of nature.

\subsection*{Acknowledgements}
We would like to thank K.~Agashe for many useful discussions and a critical reading of the manuscript.
The work of R.N.M. was supported by
the NSF grant PHY-0968854, and the work of I.S. was supported by
the U.S.~Department of Energy through grant DE-FG02-93ER-40762.
C.S.F would like to thank C.N. Yang Institute for Theoretical Physics
for the generous support.

\end{document}